\documentstyle[twocolumn,prl,aps,psfig]{revtex}                
\begin{document}
\newcommand{\tr}{\mbox{tr}}
\newcommand{\Od}{{\cal O}}
\newcommand{\Ima}{\hbox{Im}\,}
\newcommand{\Rea}{\hbox{Re}\,}
%
%
\newcommand{\NP}[1]{ Nucl.\ Phys.\ {\bf #1}}
\newcommand{\ZP}[1]{ Z.\ Phys.\ {\bf #1}}
\newcommand{\RMP}[1]{ Rev.\ of Mod.\ Phys.\ {\bf #1}}
\newcommand{\PL}[1]{ Phys.\ Lett.\ {\bf #1}}
\newcommand{\NC}[1]{Nuovo Cimento {\bf #1}}
\newcommand{\AN}[1]{Ann. Phys. {\bf #1}}
\newcommand{\PRep}[1]{Phys.\ Rep.\ {\bf #1}}
\newcommand{\PR}[1]{Phys.\ Rev.\ {\bf #1}}
\newcommand{\PRL}[1]{ Phys.\ Rev.\ Lett.\ {\bf #1}}
\newcommand{\MPL}[1]{ Mod.\ Phys.\ Lett.\ {\bf #1}}
%
%
\newcommand{\mpi}{m_\pi^2}
\newcommand{\mk}{m_k^2}
\newcommand{\me}{m_\eta^2}
\newcommand{\sq}{\mpi + \mk}
\newcommand{\fpi}{f_\pi^2}
\newcommand{\fpc}{f_\pi^4}
\newcommand{\fk}{f_K^2}
\newcommand{\fe}{f_\eta^2}
\newcommand{\fkc}{f_K^4}
\newcommand{\ep}{E_\pi}
\newcommand{\ek}{E_K}
\newcommand{\ee}{E_\eta}
\newcommand{\epp}{E'_\pi}
\newcommand{\ekp}{E'_K}
\newcommand{\eep}{E'_\eta}
\newcommand{\ct}{\cos \theta}
\newcommand{\nn}{\nonumber}
\newcommand{\spa}{\quad\quad\quad}
\newcommand{\La}{{\cal L}}
\newcommand{\oK}{\bar{K}}
\newcommand{\Kma}{K^+}
\newcommand{\Kst}{K^*}
\newcommand{\Kme}{K^-}
\newcommand{\Opd}{O(p^2)}
\newcommand{\Opc}{O(p^4)}
\newcommand{\gev}{\mbox{GeV}}
\newcommand{\cpt}{$\chi$PT}
\arraycolsep=0.5pt

\twocolumn

\title{ Non-perturbative  
Approach to effective chiral Lagrangians and
Meson Interactions }
\draft
\author{ J. A. Oller$^{1}$, E. Oset$^{2}$}
\address{Departamento de F\'{\i}sica Te\'orica and I.F.I.C.\\
Centro Mixto Universidad de Valencia - C.S.I.C.\\
46100 Burjassot (Valencia) - Spain.}

\vskip .5 cm

\author{ J. R. Pel\'aez$^{3}$}
\address{
Departamento de F\'{\i}sica Te\'orica.\\
Universidad Complutense. 28040 Madrid. Spain.}
\date{August 1997}
\maketitle
\footnotetext[1]{electronic address:oller@condor.ific.uv.es}
\footnotetext[2]{electronic address:oset@evalvx.ific.uv.es}
\footnotetext[3]{electronic address:pelaez@eucmax.sim.ucm.es}
\begin{abstract}
We develop a coupled channel unitary approach describing
the behavior at higher energies of systems
whose low-energy dynamics is given by effective
$\Opd$ and $\Opc$ chiral Lagrangians. 
Our free parameters are those of the $\Opc$ 
Lagrangian. When applied to the 
meson-meson interaction, it yields a remarkable agreement
with data up to $\sqrt{s}\simeq1.2\,\gev$, dynamically generating
the $\sigma, f_0, a_0,\rho$ and $\Kst$ resonances.
Further applications are also proposed.
\end{abstract}

The effective chiral Lagrangian formalism has become
a widespread tool to address the problem of the interaction 
of Goldstone Bosons \cite{gale,appe}. The most significant
example is Chiral Perturbation Theory ($\chi$PT) \cite{gale},
which successfully describes the low-energy behavior
of the meson-meson interaction. To lowest order, 
$\Opd$, the parameters of the chiral Lagrangian are basically masses 
and decay constants. The $\Opc$ contains several free parameters;
for instance, in \cpt, 12 chiral parameters 
are needed to describe the meson-meson 
interaction. In the case of the Standard Model 
Strongly Interacting 
Symmetry Breaking Sector (SISBS), one 
needs 13 parameters.
 The limitations of the perturbative chiral approach are obvious 
since one cannot attempt to obtain resonances and it is constrained
to low energies. Therefore
non-perturbative schemes become necessary.
This is the case when describing the meson-meson interaction up 
to about 1.2 GeV, where one finds the $\sigma$, $f_0(980)$, $a_0(980)$ in 
the scalar sector and the $\rho(770)$, $\Kst(892)$, $\phi(1020)$
 in the vector channels. The same happens if one tries to study
the expected resonance spectrum of the SISBS 
at the Large Hadron Collider.

An attempt to extend the ideas of chiral symmetry to the 
non-perturbative regime, constructing a unitary t-matrix, was done 
in ref.\cite{truongdo} using the Inverse Amplitude Method (IAM)
\cite{truong}. This approach proved efficient in reproducing low energy 
data and produced poles in the amplitudes associated to the $\rho$ and 
$\Kst$ in the vector channel as well as the $\sigma$ in the scalar channel. 
It has also been applied to study the SISBS resonances that could appear 
at LHC \cite{LHCres}.
Since only elastic unitarity
was imposed in the IAM, multichannel problems could not be
addressed. As a consequence, neither the $f_0$ and $a_0$ resonances, 
nor the inelasticities could be obtained. A similar problem with coupled channels
could also appear in the SISBS if the top quark 
couples strongly to longitudinal gauge bosons.

The treatment of coupled channels
has proved to be crucial in order to reproduce the basic features 
of the $f_0$ and $a_0$ resonances \cite{weinis}.
Another non-perturbative method, using coupled 
channel Lippmann-Schwinger 
(LS) equations was done in ref.\cite{olset}, using the $\Opd$ $\chi$PT 
amplitudes, $T_2$. A similar work in the $K-N$ system was  carried out in 
ref.\cite{kaisi}.

The LS equations used in \cite{olset} read, in matrix form,

\begin{equation}
T=T_2+\overline{T_2GT}
\label{prim}
\end{equation}
where $T_2$ and $T$ are $2 \times 2$ matrices. 
The channels of the mesonic scalar 
sector in that work are $\pi\pi$, $K \bar{K}$ for 
isospin $I=0$ as well as
$K\bar{K}$, $\pi \eta$ for  $I=1$. In eq.(\ref{prim}) 
$\overline{T_2GT}$ is given by 

\begin{equation}
(\overline{T_2GT})_{il}=i \int \frac{d^4q}{(2 \pi)^4} 
\frac{T_{2\;ij}(k,p;q)}{q^2-m^2_{1j}+i\epsilon} 
\frac{T_{jl}(q;k',p')}{(P-q)^2-m^2_{2j}+i\epsilon}
\label{VGT}
\end{equation}
with $k,p$ the four-momentum of the initial mesons and $P=p+k$. 
An important point realized in ref.\cite{olset} is that if $T_2(k,p;q)$ 
is separated in an on-shell part plus a residual term then the latter, 
when used in the loops of eq.(\ref{VGT}), does not have to be calculated,
since it only leads to coupling and mass 
renormalization. This allows the on-shell 
factorization of $T_2$ and $T$ from eq.(\ref{VGT}) 
reducing 
the LS equations to pure algebraic relations:

\begin{equation}
T=T_2+T_2 \cdot G \cdot T 
\end{equation}
Thus, we obtain
\begin{equation} 
T=[1-T_2\cdot G ]^{-1} \cdot T_2
\label{LS}
\end{equation}
where $G$ is a diagonal 
matrix given by

\begin{equation}
G_{ii}=i \int \frac{d^4 q}{(2 \pi)^4} \frac{1}{q^2-m^2_{1i}+i\epsilon}
\frac{1}{(P-q)^2-m^2_{2i}+i\epsilon}
\label{Gii}
\end{equation}

A cut-off, $q_{max}$, in the integral over $\vert \vec{q} \vert$ in eq.(\ref{Gii}) 
was adopted as a regularization method and $q_{max}$ was adjusted to 
data. With a value of $q_{max}=0.9$ GeV a remarkable agreement was obtained
for $J=0$, in the $I=0,1$ channels. However, a simple 
extrapolation of the model to the $J=1$ channel fails to reproduce the 
$\rho$ and $K^*$ resonances, which however are nicely reproduced in the 
scheme of ref.\cite{truongdo}.

It seems clear that there is important dynamics in the $\Opc$ chiral 
Lagrangian, particularly related to vector mesons, 
which cannot be generated by the LS resummation. In fact the 
resonance saturation hypothesis \cite{ega} assumes that the 
$\Opc$ terms are generated through the exchange of resonances. 

We propose here a simple and rather general 
unitary scheme in coupled 
channels which combines elements of the works
in refs.\cite{truongdo} and \cite{olset}, that can be 
now obtained as particular cases.
When applied to the meson-meson interaction, this method
describes
 simultaneously all the  
channels and reproduces the resonances below $1.2$ GeV.

With $N$ coupled channels, and for a given $(I,J)$, unitarity 
imposes (recall that $T=T^T$)
\begin{equation}
\Ima T =T \sigma T^*
\label{TsigmaT}
\end{equation}  
where $\sigma$ is an $N$-diagonal matrix 
accounting for the phase space
in each channel. 
With the normalization of \cite{Mandl},
that we adopt in the following, $\sigma=\Ima G$ of eq.(\ref{Gii}).
Thus, eq.(\ref{TsigmaT}) can be cast as
\begin{eqnarray}
\Ima G&=-\Ima T^{-1}
\end{eqnarray}

which is more conveniently rewritten as
\begin{eqnarray}
T&=&[\Rea T^{-1}-i \Ima G]^{-1}\nn\\
&=& T_2 \cdot
[T_2\cdot \Rea T^{-1} \cdot T_2- i T_2\cdot 
\Ima G\cdot T_2]^{-1}\cdot T_2
\label{method}
\end{eqnarray}

The next step is to realize that, although $T$ is 
certainly a poorly convergent function in the chiral 
expansion (above 
$500$ MeV for mesons), 
and particularly close to poles, the function $T_2\cdot T^{-1} \cdot T_2$ 
may converge much faster. The results we obtain, seem to support this
conjecture.
 (Intuitively one can imagine a function such as $T\sim \tan(x)$, 
 $T_2\sim x$ and 
$T_2\cdot T^{-1} \cdot T_2 \sim x^2\hbox{ctg}(x)$, expanded around $x=0$. 
This expansion of $T$ for values of
$x$ around $\pi/2$, where $T$ has a pole, 
is very poorly convergent, whereas $T_2\cdot T^{-1}\cdot T_2$ converges very fast).

Expanding $T$ within the chiral formalism up to $\Opc$, we have

\begin{equation}
T\simeq T_2+T_4+...\,;\,T^{-1}=
T_2^{-1}(1-T_4T_2^{-1}+...)
\label{expansions}
\end{equation}
where $T_4$ is the pure $\Opc$ $\chi$PT amplitude. Therefore
\begin{equation}
T_2\cdot \Rea T^{-1}\cdot T_2\simeq 
T_2-\Rea T_4+...
\label{VRV}
\end{equation}
The above derivation is formal since $T_2$ and $T$ 
may not be invertible. Indeed, that happens for the
$(I,J)=(1,1)$ channel. That is the reason why we
write $T$ as in eq.(\ref{method}). In that way,
we only need the expansion of $T_2\cdot \Rea T^{-1}\cdot T_2$,
which can be obtained by continuity and, as it can be seen in
eq.(\ref{VRV}), is well behaved. Finally, from eqs.(\ref{method})
and (\ref{VRV}), we arrive at
\begin{equation}
T=T_2\cdot[T_2-T_4]^{-1}\cdot T_2
\label{IAMgen}
\end{equation}
This, by itself, is an interesting result since, 
at $\Opc$, it is the coupled channel 
generalization of the IAM \cite{truongdo}.
Apart from that, this method has other advantages. Indeed, as it is
well known \cite{Penn}, the IAM with a single channel has problems
around the Adler zeros in the scalar sector. For instance, eq.(\ref{IAMgen})
with just one channel yields a double zero. The problem is actually
more serious since the expansion of $T^{-1}$ in 
eqs.(\ref{expansions},\ref{VRV}) is meaningless if $T_2=0$ or
$T_2<T_4$.
However, the expansion of $T^{-1}$ in eq.(\ref{expansions}) holds as a
matrix relation even if one of the matrix elements of $T_2$ vanishes
or is smaller than the corresponding one of $T_4$. Furthermore, the
Adler zeros appear as single zeros with the coupled channel method.

Nevertheless, eq.(\ref{IAMgen}) 
requires the complete evaluation of the 
whole $T_4$ matrix, including loops, which is rather involved.
Indeed, at present, there are only $\Opc$ calculations 
of meson-meson scattering for 
$\pi\pi$ and $\pi K$ \cite{gale,mei}. 
However, inspired in the 
LS equations, we will obtain a good approximation to 
 eq.(\ref{IAMgen}), without the
complete $\Opc$ calculation.

Let us then first reinterpret the LS equations 
discussed above using 
eqs.(\ref{method}) and (\ref{VRV}): 
It is enough to take $G$ as in eq.(\ref{Gii}),
with an appropriate cut-off such that 
$\Rea T_4 \simeq T_2\cdot \Rea G\cdot T_2$
and we recover eq.(\ref{LS}).
Note that in this way, we are accounting for the
s-channel one-loop diagram responsible for the 
unitarity logarithms. In addition, the freedom 
in the cut-off allows one to reproduce the relevant 
counterterm contribution in the meson scalar sector, 
as shown in \cite{olset}.

However, such approximations cannot hold in all meson-meson
channels, since other $\Opc$ counterterms become essential. 
Therefore, in the present work, we are also considering the
polynomial coming from the $\Opc$ tree level contributions, that we
denote by $T^P_4$, and we write

\begin{equation}
\Rea T_4 \simeq T^P_4 + T_2 \cdot \Rea G\cdot T_2 
\label{appreat4}
\end{equation}
The coefficients of $T^P_4$ 
are combinations of the $\Opc$ chiral parameters (which are usually denoted
$L_i$ within $\chi$PT and $\alpha_I$ or $a_i$ in the SISBS).

Note that $T_4^P$ is not just a vehicle for free parameters. It also
ensures that we are considering the most general $\Opc$ polynomial
structure compatible with Chiral Symmetry and its breaking. 
Once we have this chiral structure, the cut-off regularization
can be now substituted by any other scheme.

In eq.(\ref{appreat4}) one is explicitely neglecting the crossed-channel
loop contributions. Although these loops do not yield an imaginary part 
in the s-channel, they do contribute to the real part. However, in the 
s-channel these contributions have a smooth dependence on the 
energy and can be effectively reabsorbed in the $L_i$ coefficients. 

Concerning tadpoles, in the equal mass limit, 
their finite contributions can be absorbed 
in the chiral parameters. When considering different masses,
it is possible to add and subtract a finite 
tadpole-like term with an "average" mass. The added part
is absorbed in the chiral parameters. The part
not absorbed is thus of the order of
the difference between physical masses and the "average" 
mass, whose value is such that these contributions are minimized.

Then, using 
eqs.(\ref{IAMgen}) and (\ref{appreat4}) we can write
our final formula:
\begin{equation}
T\simeq T_2\cdot[T_2-T^P_4-T_2\cdot G\cdot T_2]^{-1}\cdot T_2
\end{equation}

Our purpose now is to illustrate how well the method that we have 
developed works in practice. To that aim, we will use it in a fit to 
meson-meson data, and we will denote the resulting parameters
by  $\hat L_i$ (see Table I).  
The reasons for changing their name 
are that, first, as we have already commented,
part of the effect of crossed channels terms 
and tadpoles will be reabsorbed in $\hat L_i$ and, second, that
we have chosen a cut-off regularization.  

In Fig.1 we show the results of our approach, represented 
by phase shifts of the $\pi\pi\rightarrow\pi\pi$, $\pi\pi\rightarrow K
\bar K$ and $K \pi\rightarrow K \pi$ reactions plus a mass distribution 
for the $a_0$ resonance in the $(I,J)=(1,0)$ channel.
The curves have been obtained with the fitted parameters given in Table 1.
The figure shows that the $f_0, a_0, \rho$ and $\Kst$ resonances
are very well reproduced and the phase shifts agree remarkably well
with data from threshold up to about
$\sqrt{s}\simeq 1.2\,\hbox{GeV}$, where the influence of multi-meson
channels should start to be more relevant. The $\sigma$ meson, around 500 MeV
and with a large width, as found in \cite{truongdo} and  \cite{olset}
is also reproduced here.

In summary, we have proposed a unitary scheme in
coupled channels that allows us to extend the applicability
range of the chiral Lagrangian formalism. 
We have shown how it can be implemented 
to $\Opc$  using the coefficients
of the second order Lagrangian as free parameters.
When applied to meson-meson interactions, the scheme is 
remarkably successful and reproduces
the different resonances and data up to about 1.2 GeV.

 Further applications that we can suggest
are $\gamma\gamma\rightarrow M M$ \cite{os}, form factors,
decay of particles with a pair of mesons in the final state,
meson-nucleon interactions, etc...
Within the SISBS it can also be very useful to study the 
coupling of longitudinal gauge bosons to transversal gauge 
bosons or top quarks. It also seems likely that, with a proper 
generalization, it could also be applied
to non-relativistic solid-state systems, as high-$T_c$ superconductors,
whose effective Lagrangian formulation is also
raising a great interest \cite{solid}.
The potential of the method to extend the usefulness
and advantages 
of the effective chiral Lagrangian formalism to regions
otherwise inaccessible  is certainly enormous.

\section*{Acknowledgments}

One of us, J. R. P. wishes to thank the kind hospitality
of the University of Valencia, where this work was carried out.
We would like to thank useful comments from A. Pich and M.R. Pennington. 
J. A. O. acknowledges financial support from the Generalitat Valenciana. 
This work was partially supported by CICYT under contracts PB96-0753 and 
AEN93-0776.

\newpage
\onecolumn

\begin{center}
\begin{tabular}{||c|c|c|c|c|c|c|c|c||}
\hline \hline
\rule{0.cm}{.6cm}$q_{max}=1.0$ GeV &$\hat L_1$&$\hat L_2$&
$\hat L_3$&$\hat L_4$
&$\hat L_5$&
$\hat L_7$&$2\hat L_6+\hat L_8$\\ \hline
\rule{0.cm}{.6cm}$\hbox{Our fit}$&0.5&1.0&
-3.2&-0.6&1.7&0.2&0.8\\ \hline\hline
\end{tabular}

{\bf Table 1:}{We list the values of the 
$\hat L_i$ parameters (in units of $10^{-3}$)
  obtained from the fit of our method to meson-meson 
  scattering data.}
\end{center}


\begin{figure}
\hbox{
\psfig{file=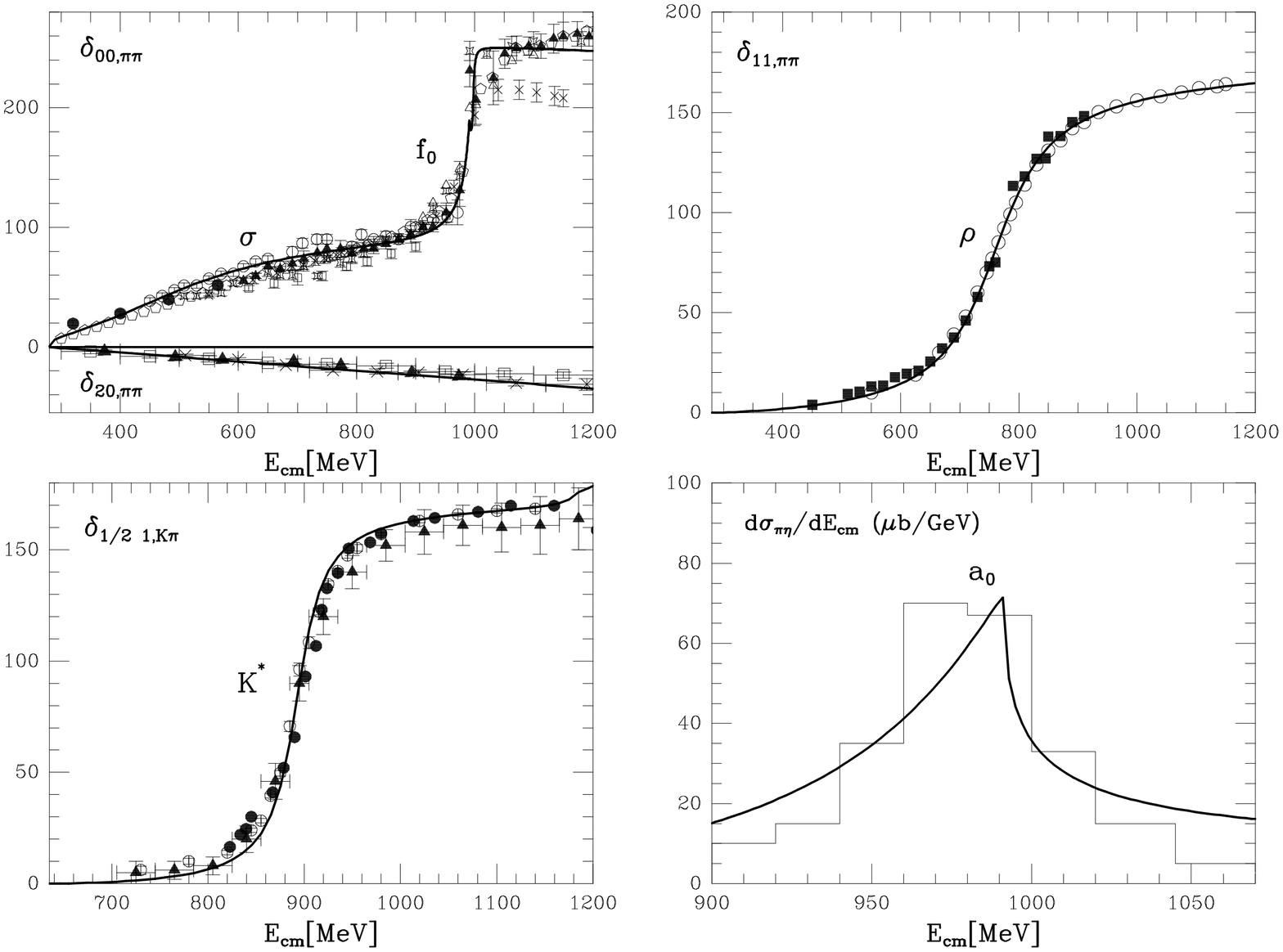,width=16cm,angle=-90}}
\hbox{
\psfig{file=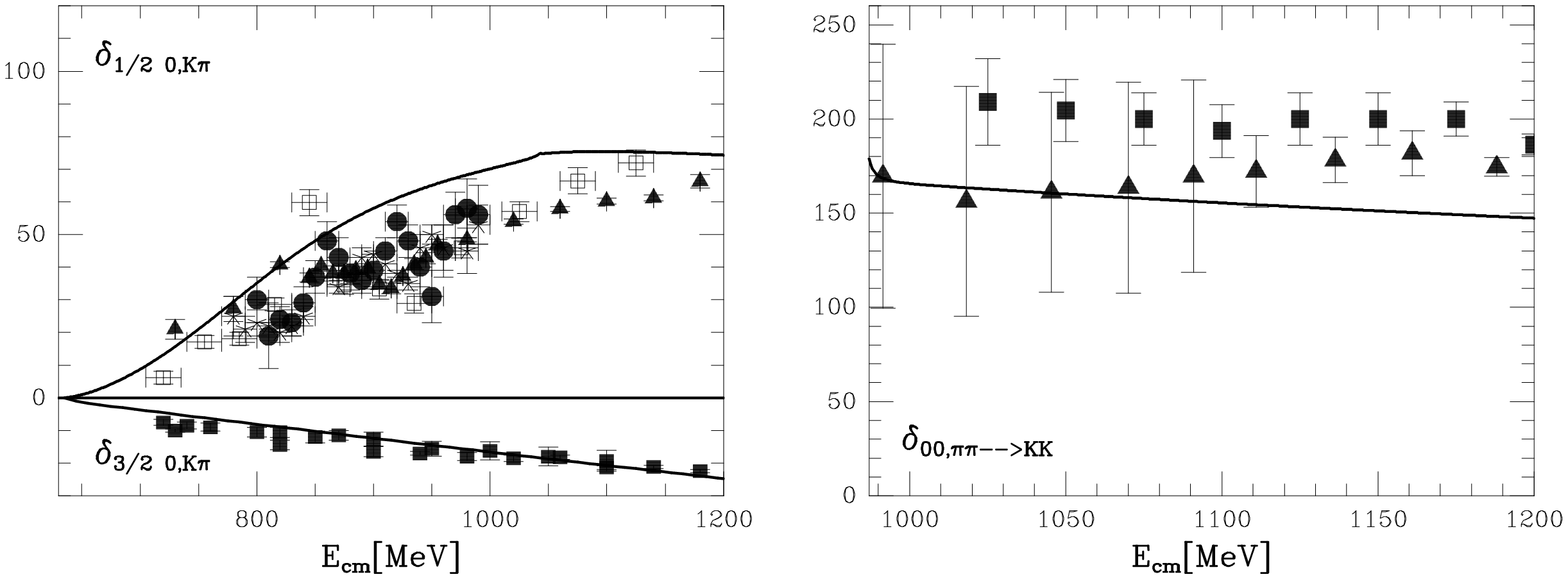,width=16cm,angle=-90}
}

{\bf Figure 1:}{We display the results of our method
for the phase shifts of $\pi\pi$
scattering in the $(I,J)=(0,0),(1,1),(2,0)$ channels, where the $\sigma$, 
$f_0$ and $\rho$ resonances appear, together with those of
$\pi\pi\rightarrow K \bar K$, as well as the phase shifts of $\pi K$
scattering in the $(3/2,0),(1/2,0) $ and $(1/2,1)$ channels, where we can see the
appearance of the $K^*$ resonance. The results also include
the $\pi^-\eta$ mass distribution 
for the $a_0$ resonance in the $(I,J)=(1,0)$ channel from
$K^-p\rightarrow \Sigma(1385)\pi^-\eta$.
For reference to the data, see \cite{truongdo} and \cite{olset}
and references therein.}
\end{figure}
\end{document}